RUB-MEP-22/94
July 1994# A haplousterotic model for the nucleon quark distribution amplitude

Robert Eckardt, Jörg Hansper, and Manfred F. Garihep-ph/9503324   14 Mar 95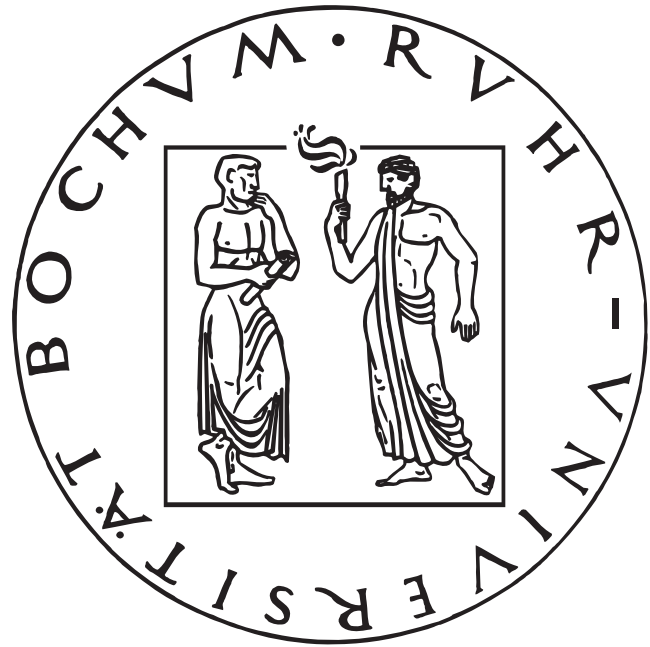

Ruhr-Universität Bochum

Institut für Theoretische Physik

Arbeitsgruppe Mittelenergiephysik

# A haplousterotic model for the nucleon quark distribution amplitude

Robert Eckardt, Jörg Hansper, Manfred F. Gari

*Arbeitsgruppe Mittelenergiephysik\*, Institut für Theoretische Physik, Ruhr-Universität Bochum,*
*D-44780 Bochum, Germany*

(July 1994)

*Using the criteria of simplicity, positivity, smoothness and process-independence of construction, we find a non-polynomial model for the quark distribution amplitude of the nucleon that is in agreement with QCD sum-rule moments. This simpler ("haplousterotic") model solves some of the problems connected with the longitudinal momentum dependence of the wave function, whereas the $\vec{k}_\perp$-dependence of the wave function, which is also important, is still left undetermined. The expansion properties of the model into a polynomial series are examined.*

The quark wave function of the nucleon is of outstanding theoretical importance because it is a universal, process-independent quantity. However, the wave function cannot be obtained *directly* by solving the equations of motion because of the complicated structure of QCD. A simplification of the problem can be made by treating the transverse momentum dependence of the wave function in an effective way so that only longitudinal momenta have to be taken into account explicitly. This is done e.g. in the definition of the so called nucleon quark distribution amplitude $\phi_N$ via the light-cone wave function $\psi_N$:

$$\phi_N(x_i, \mu^2) := \int_{\vec{k}_\perp < \mu^2} [d^2 k_{\perp i}] \, \psi_N(x_i, \vec{k}_{\perp i}) \quad . \qquad (1)$$

$\phi_N(x_1, x_2, x_3; \mu^2)$ can be interpreted as the probability amplitude for the hadron to consist of quarks collinear up to the scale $\mu^2$ with longitudinal momentum fractions $x_i$. $\psi_N(x, \vec{k}_\perp)$ is the so-called light-cone Fock-state wave function of the hadron, which has to be understood as the probability amplitude for the hadron to consist of 3 quarks with longitudinal momentum fractions $x_i$ and transverse momenta $\vec{k}_{i,\perp}$.

Some models for the quark distribution amplitude $\phi_N$ were used to construct the full wave function by multiplying it with an additional function that has also a dependence on the transverse momenta, e.g.

$$\psi_N(x, \vec{k}_\perp) = \phi_N(x) \cdot f(x, \vec{k}_\perp) \quad . \qquad (2)$$

Although the problem of constructing a realistic wave function for the nucleon in a reliable way is of outstanding interest, this task is far from being completed. On the contrary, a large number of different models for the nucleonic quark distribution amplitude has been put forward in the past by various authors. As the most important examples, we show in Fig. 1 some models for the nucleon from Refs. [2–8]. Among these, the most recent one is the "heterotic" quark distribution amplitude [8] which was constructed to amalgamate characteristic features of the preceeding models of Refs. [4,3]. We see large variations among the different wave functions which indicates that moments calculated from QCD sum rules are not sufficient in order to fix the shape of the distribution function in a reliable way.

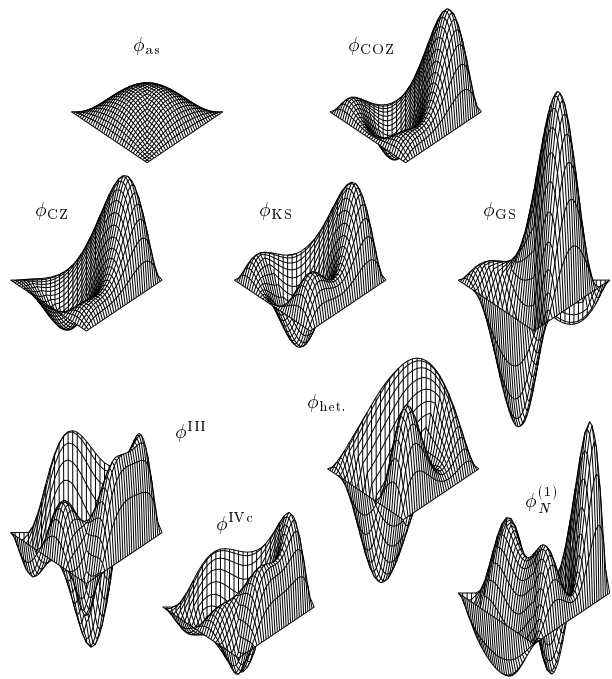

FIG. 1. Some representative models for the nucleon wave function appearing in the literature. The asymptotic form $\phi_{N,\mathrm{as}} = 120\, x_1 x_2 x_3$ is given for comparison. The different amplitudes were taken from the following references: $\phi_{\mathrm{COZ}}$: [3]; $\phi_{\mathrm{CZ}}$: [2]; $\phi_{\mathrm{KS}}$: [5]; $\phi_{\mathrm{GS}}$: [4]; $\phi_{\mathrm{het.}}$: [8]; $\phi^{\mathrm{III}}$, $\phi^{\mathrm{IVc}}$: [6]; $\phi_N^{(1)}$: [7]. All these models are polynomial models of $2^{\mathrm{nd}}$ and $3^{\mathrm{rd}}$ order. For a discussion on polynomial models see [9].

---

\*E-mail: QCD-MEP@gari.MEP.Ruhr-Uni-Bochum.de

Recently, it has been shown in Refs. [9,10] that it is not possible to find polynomial models for hadronic distribution amplitudes from their corresponding moments calculated by means of QCD sum-rules. The reason for this is the extreme instability of the polynomial expansion with respect to small variations in the moments. Already small deviations from the exact moment values result in a "neurotic" polynomial expansion, i.e. strong oscillations occur that completely distort the shape of the original function. This is discussed in detail in Ref. [10]. The experience with polynomial models indicates that oscillating wave functions are no serious candidates for the true distribution amplitude, even if additional constraints are imposed [6,8].

However, for perturbative QCD calculations of exclusive processes it is necessary to have at least some simple model for the distribution amplitude to deal with.

Consequently, one has to formulate some strong criteria that a candidate for a realistic model wave function has to fulfill *in addition to QCD sum-rule moments* so as to rule out the pathological cases [1,11–13]. An especially important criterion is the requirement of *simplicity*. This means that the functional form of the model has to be as simple as possible with a minimum number of free parameters that can be adjusted to the moments. Furthermore, the sum-rule moments of Ref. [3] are positive or allow positive values within their error bars. It is therefore sensible to require the function to be *strictly positive*. Another point is that, since polynomial expansions have turned out to be more than dangerous, the wave function has to differ *essentially* from a finite polynomial. Finally, the wave function has to be determined in a *process-independent* way, in contrast to the models in the past which were adapted to experimental values of form factors or decay widths.

We summarize the criteria for the model distribution of the nucleon that are essential in our opinion:
(i) functional simplicity, (ii) minimum number of free parameters, (iii) smoothness, (iv) absence of oscillations, (v) strict positivity, (vi) substantially non-polynomial (e.g. exponential) form and (vii) process-independent construction.

*The haplousterotic model*

According to these criteria, we look for a model wave function of simpler ("haplousterotic") type by modelling the completely symmetric asymptotic distribution amplitude, mainly by shifting the position of the maximum. We want to stress that, if this leads to a function that fulfills all requirements, there would be a convincing alternative to other wave functions that have a more complicated structure. Since we know that the method of constructing a finite polynomial expansion *directly* from sum-rule moments is inapplicable, a possible candidate for such a wave function is an exponential function that is also smooth and positive by construction. We mention that functions of such type have already been suggested by the authors of Ref. [14].

As a specific ansatz for our "haplousterotic" ($Ha^+$) model of the quark distribution amplitude of the nucleon, we choose the following form:

$$\phi_N^{Ha^+}(x) = N \exp\left(-(\frac{b_1^{(r)}}{x_1^r} + \frac{b_2^{(r)}}{x_2^r} + \frac{b_3^{(r)}}{x_3^r})\right) \quad . \quad (3)$$

The three parameters $b_i^{(r)}$ are adjusted in order to change the position of the maximum according to the requirements of QCD sum-rule moments, $r = (\frac{1}{2}, 1, 2, \ldots)$ for model (A, B, C, $\ldots$), and $N$ is the normalization factor. In other words: the $b_i^{(r)}$ are chosen so as to yield optimal agreement between the moments of the model distribution amplitude,

$$\langle ijk \rangle := \langle x_1^i x_2^j x_3^k \rangle := \int [dx]\, x_1^i\, x_2^j\, x_3^k \cdot \phi_N^{Ha^+}(x_1, x_2, x_3)\, ,$$

(4)

and the corresponding values of the same moments as obtained from QCD sum rules.

We verify that our criteria for the wave function are indeed satisfied: (i) functional simplicity: exponential function, (ii) minimum number of parameters: the $b_i^{(r)}$ are three *independent* parameters that are indispensable to determine the location of the maximum and its width, the model (iii) is smooth, (iv) has no oscillations, (v) is positive, (vi) substantially non-polynomial and (vii) no experimental information has been used to construct the model — the only input is given by QCD sum-rule moments!

For $r = 1/2$ model A reproduces the 19 moments of Ref. [3] within an accuracy of about 20 % with only *three* parameters! The corresponding moments are shown in Table I in comparison with those of $\phi_{as}$ Ref. [5] and Ref. [3]. Furthermore, also functions for different $r$ show a strong similarity with model A, if the moments are sufficiently reproduced. We have therefore another criterion for the credibility of a model constructed from QCD sum-rule moments: the shape of the resulting model function should not depend too strongly on the specific ansatz used [12].

As an example, we show plots of model A in Fig. 2. In Fig. 3 we illustrate how the position of the maximum is shifted from its (asymptotic) position in the middle of the triangle.

TABLE I. Two different sets of sum-rule moments of King and Sachrajda [5] and of Chernyak, Ogloblin and Zhitnitsky [3], in comparison with the moments of the following wave functions: $\phi_{N,\mathrm{as}}$, $\phi_{COZ}$ [3] and our new model $\phi_N^{Ha^+}$.

| $n_1\,n_2\,n_3$ | KS | COZ | | $\phi_{N,\mathrm{as}}$ | $\phi_{COZ}$ | $\phi_N^{Ha^+}$ |
|---|---|---|---|---|---|---|
| 0 0 0 | 1 | 1 | | 1 | 1 | 1 |
| 1 0 0 | 0.46 ... 0.59 | 0.560 | $^{+0.06}_{-0.02}$ | $0.33\overline{3}$ | 0.579 | 0.534 |
| 0 1 0 | 0.18 ... 0.21 | 0.192 | $^{+0.008}_{-0.012}$ | $0.33\overline{3}$ | 0.192 | 0.209 |
| 0 0 1 | 0.22 ... 0.26 | 0.229 | $^{+0.021}_{-0.029}$ | $0.33\overline{3}$ | 0.229 | 0.257 |
| 2 0 0 | 0.27 ... 0.37 | 0.350 | $^{+0.07}_{-0.03}$ | $0.\overline{142857}$ | 0.369 | 0.323 |
| 0 2 0 | 0.08 ... 0.09 | 0.084 | $^{+0.004}_{-0.019}$ | $0.\overline{142857}$ | 0.0680 | 0.0704 |
| 0 0 2 | 0.10 ... 0.12 | 0.109 | $^{+0.011}_{-0.019}$ | $0.\overline{142857}$ | 0.0890 | 0.0956 |
| 1 1 0 | 0.08 ... 0.10 | 0.090 | $^{+0.01}_{-0.01}$ | $0.\overline{095238}$ | 0.0970 | 0.0942 |
| 1 0 1 | 0.09 ... 0.11 | 0.102 | $^{+0.008}_{-0.012}$ | $0.\overline{095238}$ | 0.113 | 0.117 |
| 0 1 1 | | $-0.03$ ... 0.03 | | $0.\overline{095238}$ | 0.0270 | 0.0441 |
| 3 0 0 | | 0.236 | $^{+0.014}_{-0.026}$ | $0.0\overline{714285}$ | 0.245 | 0.212 |
| 0 3 0 | | 0.032 | $^{+0.008}_{-0.004}$ | $0.0\overline{714285}$ | 0.0381 | 0.0300 |
| 0 0 3 | | 0.052 | $^{+0.004}_{-0.004}$ | $0.0\overline{714285}$ | 0.0485 | 0.0435 |
| 2 1 0 | | 0.045 | $^{+0.004}_{-0.004}$ | $0.03\overline{5714285}$ | 0.0587 | 0.0489 |
| 2 0 1 | | 0.050 | $^{+0.005}_{-0.006}$ | $0.03\overline{5714285}$ | 0.0658 | 0.0617 |
| 1 2 0 | | 0.035 | $^{+0.002}_{-0.008}$ | $0.03\overline{5714285}$ | 0.0243 | 0.0277 |
| 1 0 2 | | 0.041 | $^{+0.002}_{-0.004}$ | $0.03\overline{5714285}$ | 0.0331 | 0.0382 |
| 0 2 1 | | $-0.004$ ... 0.007 | | $0.03\overline{5714285}$ | 0.0056 | 0.0127 |
| 0 1 2 | | $-0.005$ ... 0.008 | | $0.03\overline{5714285}$ | 0.0073 | 0.0138 |
| 1 1 1 | | — | | $0.\overline{0238095}$ | 0.0141 | 0.0176 |

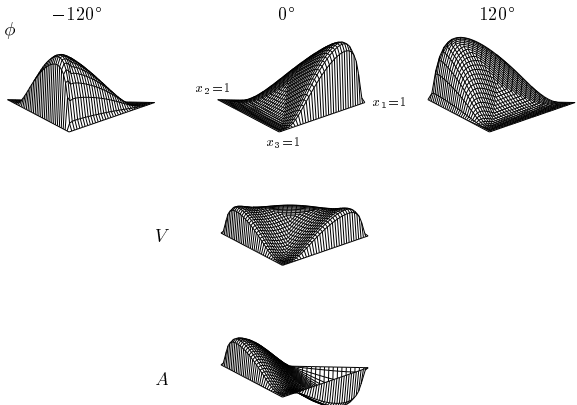

FIG. 2. The typical shape of the "$Ha^+$" (simpler) distribution amplitude determined to reproduce the QCD sum-rule moments of COZ. $\phi_N^{Ha^+}$ agrees also with the QCD sum-rule moments of KS. Note that this model function has only three free parameters, which were determined only by a best fit to the COZ moments. No attempt was made to adjust form factors or decay widths. Thus, the parameters were determined process-independently. The results do not depend very much on the precise form of the function (e.g. $r = \frac{1}{2}, 1, 2$). Although the function seems to enter the $x_2 = 1$-side very steeply, it reaches $x_i = 0$ with slope 0 – as is clearly seen on the $x_1 = 0$-side. For reference also $V_{123} = \frac{1}{2}(\phi_{213} + \phi_{123})$ and $A_{123} = \frac{1}{2}(\phi_{213} - \phi_{123})$ are given.

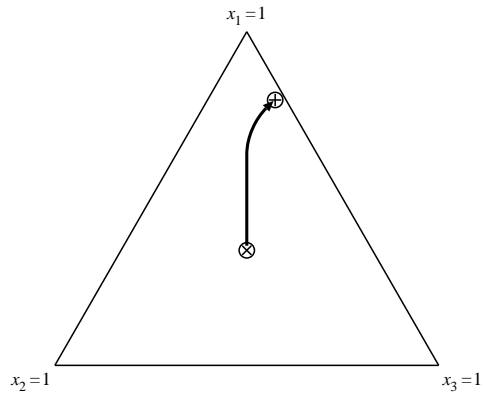

FIG. 3. This figure illustrates the deformation of the distribution amplitude away from its asymptotic form. Essentially, the maximum of the distribution amplitude moves from the completely symmetric starting point ($\otimes$) very near to the side with $x_2 = 0$ and towards $x_1 = 1$ ($\oplus$). We found for several models always $x_1^{\max} > 2/3$. This simple method suffices to reproduce the QCD sum-rule moments within a reasonable accuracy.

It seems therefore that some of the problems of constructing a (purely longitudinal) quark distribution amplitude of the nucleon have been solved: the longitudinal momentum dependence of the wave function can now be fixed sufficiently by sum-rule moments without having recourse to experimental input. It follows that the main uncertainty in the construction of the full wave function lies in the still undetermined $\vec{k}_\perp$-dependence.

*Evolution and convergence properties of the model*

Since we have now an *exact* functional form for our model distribution amplitude of the nucleon, there is in principle no reason to use a polynomial expansion in practical calculations. However, in order to obtain the correct $Q^2$-evolution of the distribution amplitude, the function has to be expanded into an infinite series of Appell polynomials $\bar{A}_i(x)$ according to

$$\phi_N^{Ha^+}(x, Q^2) = \phi_{\rm as}(x) \sum_{i=0}^{\infty} c_i \left( \frac{\alpha_S(Q^2)}{\alpha_S(\mu_0^2)} \right)^{\gamma_i} \bar{A}_i(x) \quad . \quad (5)$$

As the precise functional form of our model is known by definition, see Eq. (3), the expansion coefficients $c_i$ can be determined *directly* via the orthonormality relations of the Appell polynomials:

$$c_i = \int [dx] \, \bar{A}_i(x) \, \phi_N^{Ha^+}(x, \mu_0^2) \quad . \quad (6)$$

We note that it is not necessary to use the Appell polynomials as constructed by Lepage and Brodsky [15] if the evolution effect is ignored. $\phi_N^{Ha^+}(x, \mu_0^2)$ can then be expanded into the polynomials originally found by Appell [16]. Since the expansion is unique, it does not depend on the set of basis polynomials.

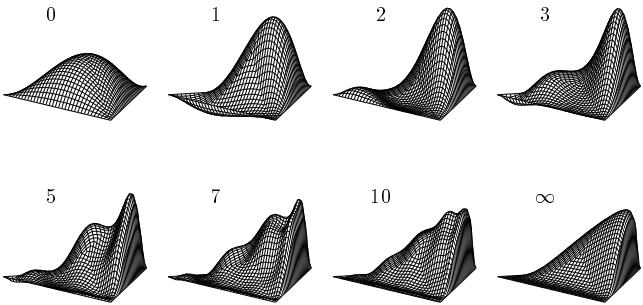

FIG. 4. The polynomial expansions of the $Ha^+$ distribution amplitude (cf. Fig. 2) in different orders. The coefficients of the Appell expansion of the best fit of $\phi_N^{Ha^+}$ are given in Table II. A polynomial expansion is needed if the evolution of the wave function is taken into account non-numerically. Although it seems that already $10^{\rm th}$ order resembles the wave function sufficiently, this is not the case for all functions of "$Ha^+$" type.

It is demonstrated in Fig. 4 that the convergence of the whole series is quite slow: a very large number of polynomials has to be summed up until the shape of the exponential function is sufficiently reproduced. We see that, although the first order gives already a qualitative impression of the rough structure of the function, the approximation is completely insufficient: the characteristic location of the maximum has moved far inside the triangle. We see only a very gradual improvement of this fact in the next orders of expansion, whereas seizable oscillations appear (we recall that we are dealing with the *exact* polynomial expansion). In seventh and higher order the expansion becomes smoother and the maximum of the distribution amplitude is located in the end-point region as required by the exact function. It seems that an expansion of order ten or more is necessary to yield an approximation of the exact function that can be used in numerical calculations. The corresponding expansion coefficients up to fifth order with respect to the normalized Appell polynomials of Lepage-Brodsky type are shown in Table II. We see that the coefficients remain of the same order of magnitude even for order 5. The slow convergence can thus be seen also numerically.

TABLE II. Expansion coefficients of the model distribution amplitude $\phi_N^{Ha^+}$ up to fifth order with respect to the Appell polynomials of Brodsky and Lepage [15]. The polynomials up tp third order are arranged according to Ref. [7] and the remaining ones according to Ref. [18]. The Appell polynomials were taken normalized to 1.

| order | $n$ | $c_n$ |
|---|---|---|
| 0 | 0 | 1 |
| 1 | 1 | 0.82916 |
|   | 2 | $-0.68548$ |
| 2 | 3 | 0.42216 |
|   | 4 | $-0.34454$ |
|   | 5 | 0.26269 |
| 3 | 6 | 0.19964 |
|   | 7 | 0.19908 |
|   | 8 | 0.17937 |
|   | 9 | 0.33544 |
| 4 | 10 | $-0.20640$ |
|   | 11 | $-0.12029$ |
|   | 12 | 0.16415 |
|   | 13 | 0.25620 |
|   | 14 | $-0.04685$ |
| 5 | 15 | 0.19552 |
|   | 16 | $-0.02693$ |
|   | 17 | $-0.13538$ |
|   | 18 | $-0.05044$ |
|   | 19 | $-0.17528$ |
|   | 20 | 0.01135 |

Summarizing our results we recall that using the criteria of simplicity, smoothness, positivity etc. one is able to construct a quark distribution amplitude which is in satisfactory aggreement with QCD sum-rule moments. We emphasize that, in view of the $\vec{k}_\perp$-dependence of the full wave function, we did not adapt the distribution amplitude to form factors or decay widths because the $\vec{k}_\perp$-dependence leaves still a large freedom in the calculation of physical observables: In the form factor calculations the end-point regions give large contributions as has been shown by Isgur and Llewellyn-Smith [17]. Therefore, the $k_\perp$-dependence of the full wave function will have a significant effect on the form factor. The effects of this $k_\perp$-dependence have been studied by means of a mass in the gluon propagator or by taking into account Sudakov form factors [19,20]. Since the resulting values are sensitive to the specific choice of the cutoff or cutoff-procedure, there is still a large range of possible values for the form factor.

Because of these reasons, we tried to determine the quark distribution in a *process-independent* way. To this end, we chose an ansatz for the model that is as simple as possible and adapted it to QCD sum-rule moments only. This model for the longitudinal momentum distibution can then be kept essentially fixed and further improvements can be attempted mainly in the construction of the correct $k_\perp$-dependence of the full wave function.

Finally, we would like to point out that because of its process-independence the quark wave function of the nucleon is an interesting object by itself, apart from perturbative QCD calculations. The calculation of form factors and other observables has nothing to do with constructing the wave function. We believe that our "haplousterotic" model is the first *simple* model of the nucleon distribution amplitude that is consistent with the requirements of QCD sum-rule moments and for which convergence is explicitly verified.

### ACKNOWLEDGMENTS

We thank J.A. Eden for useful discussions. This work was supported by the Deutsche Forschungsgemeinschaft (Ga 153-13-2).